\documentclass[prd,twocolumn,twoside,preprintnumbers,superscriptaddress,nofootinbib]{revtex4}

\usepackage{amsmath}
\usepackage{graphicx}
\usepackage{dcolumn}
\usepackage{bm}
\usepackage[hyperfootnotes=false]{hyperref}
\usepackage{xspace}
\usepackage{color}

\newcommand{\GeV}{\,\text{GeV}}
\newcommand{\TeV}{\,\text{TeV}}
\newcommand{\be}{\begin{equation}}
\newcommand{\ee}{\end{equation}}
\newcommand{\bea}{\begin{eqnarray}}
\newcommand{\eea}{\end{eqnarray}}

\begin{document}
%%%%%%%%%%%%%%%%%%%%%%%%%%%%%%%%%%

\title{The Flavour Portal to Dark Matter}

\author{Lorenzo Calibbi}
\affiliation{Service de Physique Th\'eorique, Universit\'e Libre de Bruxelles,
C.P. 225, B-1050 Brussels, Belgium}
\author{Andreas Crivellin}
\affiliation{CERN Theory Division, CH-1211 Geneva 23, Switzerland}
\author{Bryan Zald\'ivar}
\affiliation{Service de Physique Th\'eorique, Universit\'e Libre de Bruxelles,
C.P. 225, B-1050 Brussels, Belgium}

\preprint{ULB-TH/15-01, CERN-PH-TH-2015-010}

%%%%%%%%%%%%%%%%%%%%%%%%%%%%%%%%%%
%%%%%%%%%%%%%%%%%%%%%%%%%%%%%%%%%%
\begin{abstract}
We present a class of models in which dark matter (DM) is a fermionic singlet under the Standard Model (SM) gauge group but is charged under a symmetry of flavour that acts as well on the SM fermions. Interactions between DM and SM particles are mediated by the scalar fields that spontaneously break the flavour symmetry, the so-called flavons. In the case of gauged flavour symmetries, the interactions are also mediated by the flavour gauge bosons.
We first discuss the construction and the generic features of this class of models. Then a concrete example with an abelian flavour symmetry is considered. We compute the complementary constraints from the relic abundance, direct detection experiments and flavour observables, showing that wide portions of the parameter space are still viable. Other possibilities like non-abelian flavour symmetries can be analysed within the same framework.
\end{abstract}
%
%\pacs{95.35.+d, 12.39.Fe, 11.30.Rd}
%{Dark matter, Chiral Lagrangians, Chiral symmetries}

%%%%%%%%%%%%%%%%%%%%%%%%%%%%%%%%%
%%%%%%%%%%%%%%%%%%%%%%%%%%%%%%%%%
\maketitle

%%%%%%%%%%%
\section{Introduction}
\label{intro}
%%%%%%%%%%%

Establishing the nature of dark matter (DM) is one of the fundamental open problems in particle physics and cosmology. A weakly interacting massive particle (WIMP) is an excellent candidate since (for GeV to TeV scale masses) it naturally provides a relic abundance consistent with observations \cite{Ade:2013ktc}. Since WIMPs must be weakly interacting, a likely possibility is that DM is a singlet under the Standard Model (SM) gauge group.
In this case the interactions with the SM particles are transmitted by mediators, i.e.~by the ``dark sector''. Many possibilities for mediators haven been discussed in the literature, among them $Z'$ models \cite{Langacker:2008yv,Chu:2011be,Alves:2013tqa,Arcadi:2013qia}, Higgs portal models \cite{Patt:2006fw,Djouadi:2005gi,Arina:2010wv,Arina:2010an,LopezHonorez:2012kv,Djouadi:2012zc,Greljo:2013wja}, Z portal \cite{Arcadi:2014lta}, pseudoscalar mediation \cite{Boehm:2014hva,Arina:2014yna}, dark color \cite{Kitano:2004sv,Kitano:2008tk,Bai:2013xga} and models with flavoured DM \cite{Hirsch:2010ru,Boucenna:2011tj,Kile:2011mn,Batell:2011tc,Kamenik:2011nb,Agrawal:2011ze,Lopez-Honorez:2013wla,Batell:2013zwa,Agrawal:2014una,Agrawal:2014aoa}.

Flavour models aim at explaining the Yukawa structure by
introducing a flavour symmetry under which the SM fermions transform
non-trivially. After this symmetry is spontaneously broken by scalar
fields (i.e. flavons), the observed fermions masses and mixing angles
are generated as non-renormalisable operators. Froggatt and Nielsen proposed the first flavour model by introducing an abelian
$U(1)$ flavour symmetry and heavy vector-like quarks~\cite{Froggatt:1978nt}.
Later on, many other possibilities have been studied, among them
$SU(2)$ \cite{Pomarol:1995xc,Barbieri:1995uv}, $SU(3)$ \cite{King:2001uz}
as well as discrete flavour symmetries like $A_4$ \cite{Altarelli:2005yp}.

In this article we propose a new mechanism generating the interactions of DM with the visible sector: DM and the SM particles are charged under a flavour symmetry and interact with each other only via flavour interactions.\footnote{Note that our framework shares this starting assumption with some
flavoured DM models, e.g. \cite{Kile:2011mn,Kamenik:2011nb,Lopez-Honorez:2013wla,Agrawal:2014aoa}. 
Nevertheless, the crucial difference is that we propose either flavour-breaking scalars or flavour
gauge bosons as the only mediators between the dark and the SM sectors.}
This is an appealing possibility since it is minimal in the sense that no {\it ad hoc} quantum numbers must be introduced and DM interactions are described by the same dynamics generating fermion masses in the SM.
Furthermore, relating Dark Matter and flavour models provides a handle on the otherwise unspecified flavour-breaking scale, motivating the possibility of low-energy realisations with interesting phenomenology 
that would allow insights into the flavour sector \cite{Calibbi:2012at}.

%%%%%%%%%%%%%%%%%%
\section{General setup}
\label{sec:gen}
%%%%%%%%%%%%%%%%%%

In this section, we examine the generic features of ``flavour portal'' models that do not depend on the specific implementation, e.g.~the transformation properties of SM and DM fields under the flavour symmetry.

In models \`a la Froggatt-Nielsen~\cite{Froggatt:1978nt,Leurer:1992wg,Leurer:1993gy,Ibanez:1994ig,Binetruy:1994ru}, 
a new symmetry of flavour ($\mathcal{G}_F$), under which the SM fermions are charged, is introduced, such that the Yukawa couplings of the SM are forbidden at the renormalisable level (with the possible exception of the top Yukawa) and fermion masses arise once the symmetry is broken as higher dimensional operators:
\begin{equation}
(m_f)_{ij}  = a^f_{ij} \left(\frac{\langle \phi \rangle}{M}\right)^{n^f_{ij}} \frac{v}{\sqrt{2}}\,.
\label{eq:yukawa}
\end{equation}
Here $v\cong 246\;{\rm GeV}$ is the vev of the SM Higgs $h$ and $\phi$ represents one or more scalar fields, the flavons, whose vev breaks the flavour symmetry. $M$ is a cutoff that can be interpreted as the mass of vector-like fermions, that constitute the UV completion of the model.\footnote{In Ref.~\cite{Calibbi:2012yj} it has been discussed that heavy scalars with the quantum numbers of the SM Higgs field can also be used for a similar mechanism. Model building details and phenomenology of the messenger sector have been discussed in Ref.~\cite{Calibbi:2012yj,Calibbi:2012at}.}
The exponents $n^f_{ij}$ are dictated by the transformation properties of the fields under $\mathcal{G}_F$, while $a^f_{ij}$
are unknown coefficients originating from the fundamental couplings of SM fields, flavons
and messengers in the UV-complete theory. They are commonly assumed to be $\mathcal{O}(1)$,
so that the the observed hierarchies of the Yukawas are solely accounted for by the flavour symmetry breaking.
This can be done by a suitable choice of the flavour charges or representations of the SM fermions 
(one can assume that the Higgs field is neutral under $\mathcal{G}_F$), provided that $\epsilon \equiv {\langle \phi \rangle}/{M} $
is a small expansion parameter, typically of the order of the Cabibbo angle or smaller.

\subsection{$\mathcal{G}_F$ as a global symmetry}
If the flavour symmetry is global, there are no extra gauge bosons associated to it, so that for a DM particle which carries flavour charge, the only interactions with the visible sector are through flavon fields $\phi$. 
In particular, we are going to consider two chiral fermions $\chi_L$ and $\chi_R$, which are SM singlets but transform non-trivially
under $\mathcal{G}_F$. In perfect analogy with the SM fermions, the DM particle acquires a Dirac mass term though the flavour symmetry
breaking:
\begin{equation}
m_\chi  = b_{\chi} \left(\frac{\langle \phi \rangle}{M}\right)^{n^\chi} \langle \phi \rangle\,, 
\label{eq:chimass}
\end{equation}
where again $b_{\chi}$ is an $\mathcal{O}(1)$ coefficient and $n^\chi$ depends on the flavour representations/charges of $\chi_L$ and $\chi_R$. Note that in this setup the stability of the DM particles is automatically guarantied (at least for what concerns the quark sector) by an accidental symmetry related to their singlet nature. 
In fact, due to Lorentz and $SU(3)$ invariance, DM could only decay by violating baryon number. However, like in the SM, baryon number is an accidental symmetry of our model, since the dark sector, i.e.~the DM fields and the DM messengers which constitute the UV completion of Eq.~(\ref{eq:chimass}), is composed only by SM singlets. This guarantees the stability of DM at least at the perturbative level, since the dark sector fields can not mix with quarks and Froggatt-Nielsen messengers (i.e.~heavy vector-like quarks) because of colour and charge conservation.

From Eqs.~(\ref{eq:yukawa}, \ref{eq:chimass}) we can obtain the couplings of both DM and SM fermions to a dynamical flavon, without further specifying the details of the model:
\begin{align}
\mathcal{L} &\supset n^f  \frac{m_f}{\langle \phi \rangle} f_L f_R \phi + (n^\chi+1) \frac{m_\chi}{\langle \phi \rangle} \chi_L \chi_R \phi
\nonumber \\
& \equiv   \lambda_f f_L f_R  \phi + \lambda_\chi \chi_L \chi_R \phi\,.
\label{eq:lagrangian}
\end{align}
Here, we suppressed flavour indices but one has to keep in mind that the couplings to fermions are in general flavour changing. In fact, $\lambda_f $ is not diagonal in the same basis as the fermion mass matrix $m_f$, as a consequence of the flavour dependence of the exponents $n^f$. The above expressions can be easily generalised to the case of multiple flavons.
%%%%%
\begin{figure}
\centering
\includegraphics[width=0.5\textwidth]{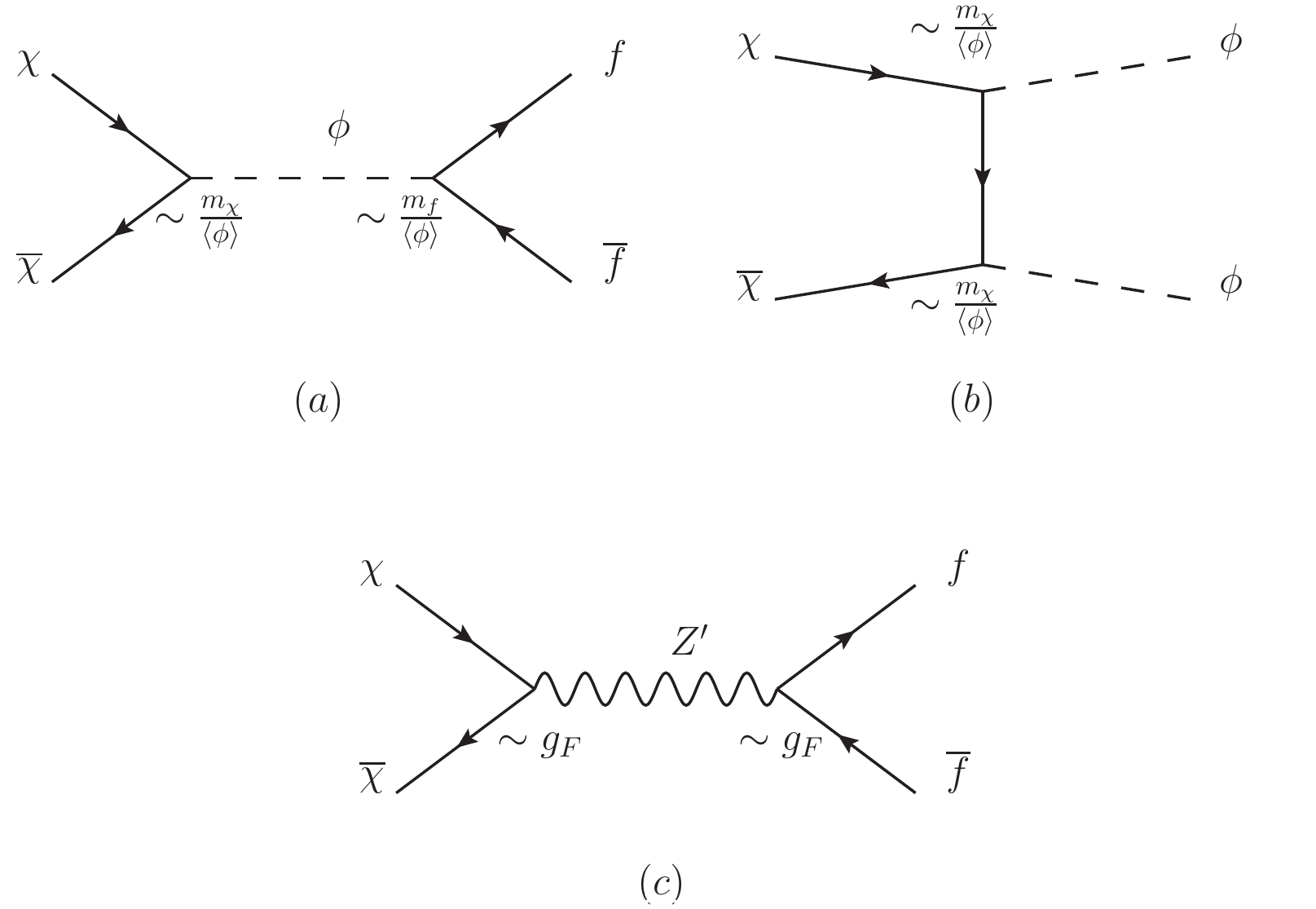}%
\vspace{-1ex}
\caption{Feynman diagrams contributing to DM annihilation in presence of light flavons ({\it a}) and ({\it b}) and light flavour gauge bosons ({\it c}).}
\label{fig:diagrams}
\end{figure}
%%%%

For what concerns the flavon mass, we assume that it also arises from the flavour symmetry breaking, such that: 
\be m_\phi = k \langle \phi \rangle\,. \ee 
Since we are not going to specify the details of the scalar potential and the symmetry breaking, 
we take $k$ as a free parameter.

From the above Lagrangian, we can see that our setup shares some similarities to the Higgs-portal scenario \cite{Patt:2006fw,Djouadi:2005gi,LopezHonorez:2012kv,Djouadi:2012zc,Greljo:2013wja}, in particular
the fact that the mediator prefers to couple to heavier fermions, but also the possible correlations between the DM annihilation and 
scattering with nuclei.

In Fig.~\ref{fig:diagrams} we show the Feynman diagrams ({\it a}) and ({\it b}) contributing to DM annihilation
in presence of light flavons. Notice that the contribution ({\it b}) requires $m_\chi > m_\phi$. 
From the figure, we can immediately infer the parametric dependence of the annihilation processes we are interested in.
Concerning the s-channel processes ({\it a}), we see that the thermally-averaged annihilation cross section mediated by flavons scales as:
\begin{align}
\langle\sigma^{\rm S}_\phi v\rangle&\sim  \frac{\lambda^2_\chi  \lambda^2_f m_\chi}{(4m_\chi^2-m_\phi^2)^2 + \Gamma_\phi^2 m_\phi^2} T\,, 
\label{sigmav1}
\end{align}
where the flavon decay width $\Gamma_\phi$ depends on $\lambda_f^2$ and $\lambda_\chi^2$.
From Eq.~(\ref{eq:lagrangian}), we see that the flavon-mediated process preferably involves the heaviest fermions that are kinematically accessible and it can be doubly suppressed in the case of heavy flavons: both by the propagator and by the couplings (that scale as $1/m_\phi$). This is the reason why, as we will see in the explicit example of the next sections, the correct relic abundance might require a resonant enhancement of the annihilation cross section, i.e.~$m_\chi\simeq m_\phi/2$. 
Note also that the annihilation cross section in Eq.~(\ref{sigmav1}) depends linearly of the temperature, which is a consequence of the velocity-suppression of the process. The contribution of the diagram ({\it b}) of Fig.~\ref{fig:diagrams} scales like
\be \langle\sigma^{\rm t}_\phi v\rangle \sim  \frac{\lambda_\chi^4}{m^3_\chi} T \ee
in the limit $m_\chi \gg m_\phi$, and features a p-wave suppression as well.

Elastic DM-nuclei scattering can only be mediated by $\phi$, hence being controlled by the same couplings depicted 
in the diagram ({\it a}) of Fig.~\ref{fig:diagrams}.
Therefore, the parametric dependence of the spin-independent cross section is:
\be
\sigma^{\rm SI}_{\phi} \sim \frac{\lambda_\chi^2\lambda_{\phi N}^2}{m^4_\phi}\mu^2_{\chi N}\,,
\label{sSI}
\ee
Here $\mu_{\chi N}$ is the DM-nucleon reduced mass, and $\lambda_{\phi N}\propto \lambda_f$ is
the scalar couplings to nucleons \cite{Crivellin:2013ipa}. 

\subsection{\bf $\mathcal{G}_F$ as a local symmetry}
We also consider the possibility that the flavour symmetry is gauged, leading to extra gauge boson(s) with coupling $g_F$. The phenomenology of the model drastically changes. The reason is that the couplings of fermions and dark matter to $Z'$ are completely determined by the flavour charges ($\mathcal{Q}_X$ in our abelian example) of the particles:
\begin{align}
\mathcal{L} \supset~ & 
 g_F\,\overline{\chi}\gamma^\mu (\mathcal{Q}_{\chi_L} P_L + \mathcal{Q}_{\chi_R}P_R)\chi~Z'_\mu+\nonumber\\
 & g_F\,\overline{f} \gamma^\mu (\mathcal{Q}_{f_L} P_L + \mathcal{Q}_{f_R}P_R) f ~Z'_\mu\,.
 \label{eq:Zpcoupl}
\end{align}
As we can see, the couplings are not suppressed by the flavour-breaking scale, as in the case of flavon mediation. 
Furthermore, unlike flavons, flavour gauge bosons do not preferably couple to heavy flavours
and the DM annihilation, depicted in Fig.~\ref{fig:diagrams} ({\it c}), is potentially efficient even for flavour-breaking scales well above the TeV. 
In fact, the flavour gauge bosons can be substantially lighter than flavons if the coupling is weak. In fact, in the abelian example above we have $m_{Z'}=\sqrt{2} g_F \langle\phi\rangle$. Of course, too small values of $g_F$ would suppress the annihilation even if $Z'$ is light. From the diagram ({\it c}) of Fig.~\ref{fig:diagrams}, we see that:
\begin{align}
\langle\sigma_{Z'} v\rangle& \sim  \frac{g_F^4}{(m_{Z'}^2-4m_\chi^2)^2 + \Gamma_{Z'}^2 m_{Z'}^2}  m_\chi^2, 
\end{align}
where the $Z'$ width $\Gamma_{Z'}$  is proportional to $g_F^2$. Also, as we can see the s-channel annihilation is not p-wave suppressed (unlike the flavon-mediation case).

As in the flavon case the $Z'$ interactions are in general flavour-violating, since quarks of different families couple in general differently to the $Z'$. This feature is going to induce severe constraints on the $Z'$ mass from flavour observables. 
The direct detection interaction is also determined by the gauge boson exchange:
\be
\sigma^{\rm SI}_{Z'} \sim \frac{g_F^2 \lambda_{Z'N}^2}{m^4_{Z'}}\mu^2_{\chi N}\,.
\ee
The vector coupling to the nucleons is $\lambda_{Z' N}\propto g_F$. 

\section{Construction of an explicit model}
\label{sec:model}

Let us now consider an explicit example of our general idea that DM is communicating with the SM fields via flavour interactions. For simplicity we choose a Froggatt-Nielsen $U(1)_F$ model.\footnote{In the case the symmetry is global, we should rather consider a discrete subgroup $Z_N \subset U(1)$, in order to avoid massless Nambu-Goldstone bosons. For large enough $N$, this does not substantially modify the effective theory, such that we can still work in terms of charges of a continuous $U(1)$ \cite{Leurer:1993gy}.}
We assign flavour charges to the SM quarks,  $\mathcal{Q}_{q_i}$, $\mathcal{Q}_{u_i}$ and $\mathcal{Q}_{d_i}$. 
The SM scalar doublet is neutral under $U(1)_F$ and the flavour symmetry is broken by a single flavon with $\mathcal{Q}_\phi = -1$. In addition we introduce new fermion which are singlets under the SM gauge group but carry $U(1)_F$ charges: the dark matter particles.

The SM Yukawas read:
\be
y^u_{ij} = a^u_{ij} \epsilon^{\mathcal{Q}_{q_i}+\mathcal{Q}_{u_j}},\quad y^d_{ij} = a^d_{ij} \epsilon^{\mathcal{Q}_{q_i}+\mathcal{Q}_{d_j}},
\label{eq:yuk}
\ee
where $a^{(u,d)}_{ij}$ are assumed to be $\mathcal{O}(1)$ numbers and $\epsilon\equiv \langle\phi \rangle/M$. We take $\epsilon = 0.2$ to reproduce the observed quarks hierarchies and define the rotations to the fermion mass basis as
\be
(V^u_L)^\dag y^u V^u_R \equiv \hat{y}^u,\quad (V^d_L)^\dag y^d V^d_R \equiv \hat{y}^d\,,
\label{eq:rot}
\ee
where $\hat{y}^{(u,d)}$ are diagonal matrices.

Given the hierarchical structure of the Yukawas, the rotations are approximated by ratios of the Yukawa entries:
\begin{align}
\label{eq:rotu}
(V^{(u,d)}_L)_{ij} &\approx  \frac{y^{(u,d)}_{ij}}{y^{(u,d)}_{jj}}= \frac{a^{(u,d)}_{ij}}{a^{(u,d)}_{jj}} \epsilon^{\mathcal{Q}_{q_i}-\mathcal{Q}_{q_j}},\\
(V^{(u,d)}_R)_{ij} &\approx  \frac{y^{(u,d)}_{ji}}{y^{(u,d)}_{jj}}= \frac{a^{(u,d)}_{ji}}{a^{(u,d)}_{jj}} \epsilon^{\mathcal{Q}_{(u_i,d_i)}-\mathcal{Q}_{(u_j,d_j)}},
\label{eq:rotd}
\end{align}
where $i\geq j$.
From Eq.~(\ref{eq:yuk}), it follows that the couplings to a dynamical flavon defined in Eq.~(\ref{eq:lagrangian}) are given by
\begin{align}
\lambda^{(u,d)}_{ij} &= a^{(u,d)}_{ij} (\mathcal{Q}_{q_i}+\mathcal{Q}_{(u_j,d_j)}) \epsilon^{\mathcal{Q}_{q_i}+\mathcal{Q}_{(u_j,d_j)}}\frac{v}{\langle\phi\rangle}\,,
\label{eq:ld}
\end{align}
Using the above expressions for the rotations, we can then easily estimate the couplings $\lambda^{(u,d)}$ in terms of fermion masses and mixing angles. For instance, we have:
\begin{align}
\lambda^u_{12}&\approx(\mathcal{Q}_{q_1}+\mathcal{Q}_{u_2})  {m_c } {(V^u_L)_{12}}/{{\langle\phi\rangle}},\\
\lambda^u_{21}&\approx(\mathcal{Q}_{q_2}+\mathcal{Q}_{u_1})  {m_c } {(V^u_R)_{12}}/{{\langle\phi\rangle}}.
\end{align}
As we can see, these couplings are thus expressed in terms of physical observables with the residual uncertainty from the unknown $\mathcal{O}(1)$ coefficients encoded in the $V_L$ and $V_R$ rotations.\footnote{Given the hierarchical structure of the Yukawa, the order of magnitude of the entries of $\lambda^{(u,d)}$ do not change after rotating to the mass basis, only the unknown  $\mathcal{O}(1)$s are modified.}
The DM mass arises as in Eq.~(\ref{eq:chimass}) with the exponent given by:
\be n_\chi =\mathcal{Q}_{\chi_L}+\mathcal{Q}_{\chi_R} -1. \ee

Therefore, the model with the global $U(1)_F$ is completely defined -- up to ${\cal O}(1)$ coefficients -- 
by the parameters: 
\be m_\phi,\,m_\chi,\, k\equiv m_\phi/\langle \phi\rangle,\ee
where a given value of $m_\chi$ can be obtained by suitable choices of $\mathcal{Q}_{\chi_L}$, $\mathcal{Q}_{\chi_R}$ and $b_\chi$.

There are few possibilities how $U(1)_F$ can be chosen such that the measured fermion masses and mixing can be reproduced. Out of the possibilities outlined in \cite{Chankowski:2005qp}, here we adopt the following example:
\begin{align}
(\mathcal{Q}_{q_1},\,\mathcal{Q}_{q_2},\,\mathcal{Q}_{q_3})&=(3,\,2,\,0),\nonumber\\
(\mathcal{Q}_{u_1},\,\mathcal{Q}_{u_2},\,\mathcal{Q}_{u_3})&=(3,\,2,\,0),\nonumber\\
(\mathcal{Q}_{d_1},\,\mathcal{Q}_{d_2},\,\mathcal{Q}_{d_3})&=(4,\,2,\,2).
\end{align}
Note that, since for us $\mathcal{Q}_{q_3}=\mathcal{Q}_{u_3}=0$, we have $\lambda^u_{33}=0$, i.e.~no unsuppressed coupling of the flavon to the top. As a consequence the largest coupling of the flavon is flavour violating: $\phi\,t_{L,R}\,c_{R,L}$. 

In the case $U(1)_F$ is local, we additionally have a $Z'$ whose couplings are shown in Eq.~(\ref{eq:Zpcoupl}). Given the charge assignment above, couplings to light generations are larger. Furthermore, once the rotations in Eq.~(\ref{eq:rot}) are applied to go to the fermion mass basis, flavour-violating couplings arise from Eq.~(\ref{eq:Zpcoupl}), of the form $\sim (\mathcal{Q}_{q_j}- \mathcal{Q}_{q_i}) \times(V^{(u,d)}_L)_{ij}$. This induces $Z'$-mediated FCNC at tree-level, setting strong limits on the gauge coupling $g_F$ for a given $Z'$ mass.

The above adopted charge assignment is anomalous \cite{Chankowski:2005qp}, as it is a common feature of 
$U(1)_F$ models that successfully reproduce the fermion masses and mixing \cite{Ibanez:1994ig,Binetruy:1994ru,Dudas:1995yu,Binetruy:1996xk}.\footnote{However, cf.~an anomaly-free solution presented in \cite{Dudas:1995yu} and the recent study \cite{Tavartkiladze:2011ex} in the context
of $SU(5)$ GUT.}
As the usually invoked Green-Schwarz mechanism would require a flavour-breaking scale close to the Planck scale, we have to assume that anomalies are canceled by the unspecified field content of the hidden and/or the messenger sector. In particular, if the flavour messengers are heavy quarks in vector-like representations of the SM gauge group, they do not need necessary to be vector-like under the flavour symmetry too, leaving large freedom to cope with the $U(1)_F$ anomalies.

%%%%%%%%%%%%%%%%%
\section{Phenomenological Analysis}
\label{Pheno}
%%%%%%%%%%%%%%%%%

\subsection{Flavour Constraints}
As we have seen, both in the global and in the local version of our $U(1)_F$ example FCNC appear at tree-level. 
Using the bounds on FCNC operators reported in \cite{Isidori:2010kg,Calibbi:2012at}, we can estimate the limits on the flavon/$Z'$ mass.

The strongest limit to the flavon mass comes from the $(\overline{s}_L d_R)( \overline{s}_R d_L)$
operator, contributing to $K-\overline K$ mixing. Integrating out the flavon the above operator is induced with the coefficient
$\lambda^d_{12}\lambda^d_{21}/m_\phi^2$. The bounds result:
\begin{align}
\label{eq:fcnc-phi}
&\Delta M_K:~m_\phi \gtrsim \sqrt{k}\times 580\,\GeV\,,\\ 
&\epsilon_K:~m_\phi \gtrsim \sqrt{k}\times \sqrt{\arg(\lambda^d_{12}\lambda^d_{21})} \times2.3\,\TeV\,,
\label{eq:epsK-phi}
\end{align}
where we neglected an overall coefficient, product of fundamental $\mathcal{O}(1)$  Yukawa-like couplings,
which can still conspire to relax the bound to some extent. 
Although the limit from CP violation in $K-\overline K$ mixing is rather strong, we see 
that a mild suppression of the overall phase, $\arg(\lambda^d_{12}\lambda^d_{21})\approx 0.1$,
is enough to reduce it at the level of the bound from $\Delta M_K$.

In the $Z'$ case, the strongest bounds also come from $K-\overline K$ 
mixing. The leading operator is 
$(\overline{s}_L \gamma^\mu d_L)( \overline{s}_R \gamma_\mu d_R)$,
whose Wilson coefficient reads:
\begin{align}
\frac{g_F^2}{m^2_{Z'}}  \Delta\mathcal{Q}_{q_1 q_2} (V^d_L)_{12} \Delta\mathcal{Q}_{d_1 d_2} (V^d_R)_{12}\,,
\end{align}
where $\Delta\mathcal{Q}_{f_1 f_2}\equiv \mathcal{Q}_{f_1}-\mathcal{Q}_{f_2}$.
The bounds on the $Z'$ mass then result:
\begin{align}
\Delta M_K:~m_{Z'} \gtrsim& \left(\frac{g_F}{10^{-3} } \right)\times 210\,\GeV\,,\\
\epsilon_K:~m_{Z'} \gtrsim& \left(\frac{g_F}{10^{-3} } \right)
\times \sqrt{\arg \left(  (V^d_L)_{12}  (V^d_R)_{12} \right) } \nonumber\\ 
& \times 3.3\,\TeV\,, 
\label{eq:fcnc-zp}
\end{align}
where again we omitted an $\mathcal{O}(1)$  uncertainty due to the coefficients entering the rotations 
in Eqs.~(\ref{eq:rotu}, \ref{eq:rotd}).
The stringent bound from $\epsilon_K$ can be relaxed at the level of the CP conserving limit if 
$\arg\left((V^d_L)_{12}(V^d_R)_{12}\right) \approx 0.01$.

The fields that constitute the UV completion of Froggatt-Nielsen models (vector-like quarks or heavy scalars) can give contributions to the FCNC that are larger than those mediated by flavon exchanges \cite{Calibbi:2012at}, even though they enter at the one-loop level. The reason is that the flavon couplings are proportional to the fermion masses, suppressing processes involving light generations. Adopting the model-independent approach of \cite{Calibbi:2012at}, we find that, in our model, the strongest bound in the hypothesis of suppressed phases to the messenger scale comes from  $D-\overline D$ mixing and it can be translated to a limit on the flavon mass:
\be m_\phi \gtrsim C_D\times k\times 2.3~\TeV,\label{eq:mess-bound}\ee 
where $C_D$ parameterises a product of unknown $\mathcal{O}(1)$ coefficients.
In presence of  $\mathcal{O}(1)$ CP-violating phases, 
we obtain the following bound from $\epsilon_K$:
\be m_\phi \gtrsim C_K\times k\times 27~\TeV.\label{eq:mess-bound-epsK}\ee 
As it will be clear from the discussion in the following subsection, the phenomenologically interesting region 
of the parameter space would be excluded by this limit, unless we assume that the overall phase in $C_K$ 
can reduce it by about one order of magnitude.

%%%%%%%%%%%%%%%%%%%%%%%%%%%%
\begin{figure*}[t]
\centering
\includegraphics[width=0.31\textwidth]{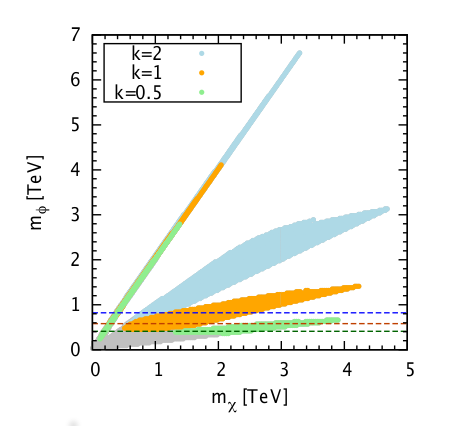}
\includegraphics[width=0.31\textwidth]{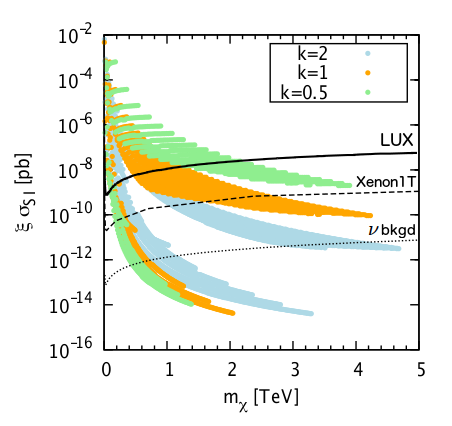}
\hspace{0.25cm}
\includegraphics[width=0.29\textwidth]{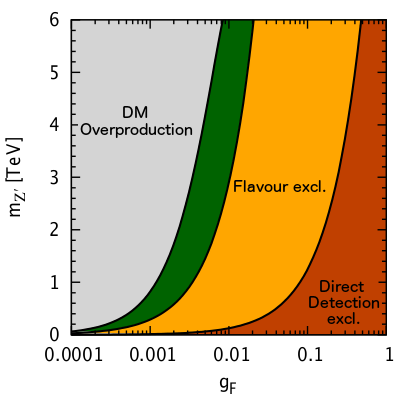}
\caption{
{\it Left}: points with $\Omega_{\rm DM} h^2 \le0.13$ in the  ($m_\chi$, $m_\phi$) plane for different values of $k$. The dashed lines represents the corresponding lower bounds on $m_\phi$ from FCNC constraints. 
{\it Centre}: nucleon-DM scattering cross section scaled by the actual DM density $\xi \cdot \sigma_{\rm SI}$  (with $\xi\equiv \Omega_\chi h^2 / 0.11$) for the same points as before.
{\it Right}: ($g_F$, $M_{Z'}$) plane in the local $U(1)_F$ case; only the green band is allowed by all data.}
\label{fig:zprime}
\end{figure*}
%%%%%%%%%%%%%%%%%%%%%%%%%%%%%

%%%%%%%%%%%%%%%%%%
\subsection{Relic abundance and Direct Detection}
%%%%%%%%%%%%%%%%%%%%%%%%%%%%

We have implemented our model in the {\tt MicrOmegas} code \cite{Belanger:2013oya} in order to obtain an accurate numerical calculation of the relic abundance and the direct detection cross section. In the case of a global $U(1)_F$, we work with the free parameters $m_\phi,\,m_\chi,\,k\equiv m_\phi/\langle \phi\rangle$ and set the ${\cal O}(1)$ coefficients to unity. 
In the gauged case, we have in addition $m_{Z'}$ and $g_F$ and the dependence 
on unknown coefficients (in the fermion rotation matrices) is much milder. We require the relic abundance not to overshoot the Planck measurement \cite{Ade:2013ktc}, taking as a conservative limit $\Omega_{\rm DM} h^2 \le0.13$.  

{\bf Global $U(1)_F$ case.}
The results for the global case are shown in the first plot of Fig.~\ref{fig:zprime}. As we can see, besides the flavon resonance, there is a wide region with $m_\chi > m_\phi$ where efficient annihilation is provided by diagram ({\it b}) of Fig.~\ref{fig:diagrams} and FCNC constraints are satisfied. 
We display here only the bound of Eq.~(\ref{eq:fcnc-phi}) from  flavon exchange, 
assuming a mild suppression of the CP-violating phases at the level discussed in the previous subsection.
We neglect the more stringent bounds of Eq.~(\ref{eq:mess-bound}) as they rely on the additional assumption of a weakly-coupled messenger sector at the scale $M$. We just notice that they would  exclude the region where $\chi\overline\chi\to\phi\phi$ provides the correct relic density,  
leaving the resonance at $m_\chi \approx m_\phi/2$ as the only viable solution, unless a mild suppression comes 
from the coefficient $C_D$. This would be also the effect of the bound from $\epsilon_K$, cf.~Eq.~(\ref{eq:epsK-phi}),
in presence of ${\cal O}(1)$ phases.
The points shaded in grey at low values of $m_\phi$ do not fulfil the LUX constraints \cite{Akerib:2013tjd}, shown in the second plot.
As we can see, most of the parameter space has good prospects to be tested by next generation direct searches experiments and the only points that might be hidden under the neutrino background correspond to resonant annihilating DM with $m_\chi \gtrsim 1$ TeV.

{\bf Local $U(1)_F$ case.}
In the local case, we find that the only viable possibility to fulfil the relic density bounds relies on resonant $Z'$ exchange, $m_\chi \approx m_{Z'}/2$, as a consequence of the stringent FCNC constraint of Eq.~(\ref{eq:fcnc-zp}).
Hence, we varied $g_F$ and $m_{Z'}$ and scanned $m_\chi$ around the resonant condition. 
The result is shown in the right plot of Fig.~\ref{fig:zprime}. We see that for a given $m_{Z'}$, too small values of $g_F$
cannot give the correct relic density, even on the resonance, while there is an upper bound on $g_F$ from the 
FCNC bound of Eq.~(\ref{eq:fcnc-zp}) and the limit given by direct detection experiments is comparably much weaker. As a result, only a narrow region, depicted in green, is still viable. The best way to probe this surviving region seems to rely on an increased sensitivity of FCNC constraints, especially in the $D-\overline{D}$ and $K-\overline{K}$ systems.

\subsection{Collider signals}
There are in principle very interesting signals that can be searched for at the LHC. In fact, the largest branching ratio of the flavon is $\phi\to t\overline c,\,\overline t c$, which is around 60\%, but also the $b\overline b$ or $b\overline s$ channels are sizeable (while the BR to $\chi\overline\chi$ is very suppressed because it is at threshold or below due to relic density constraints), leading for example to final states like $pp\to t\overline c t\overline c$ or $pp\to b\overline bb\overline b$. However, the production cross section of $\phi$ is small, at the level of $0.1 (10^{-3})$ fb for LHC$@$14 TeV for flavons with $m_\phi = 500~(1000)$ GeV, values already at the edge of the flavour constraints. A similar situation happens in the $Z'$ case, where flavour bounds make any LHC signal very challenging.  

Finally, let us mention that Higgs-flavon mixing can be induced by quartic terms in the scalar potential such as $H^\dag H \phi^\dag \phi$ as well as by loops of SM fermions. We checked that 
in the former case the mixing scales like $(v/\langle\phi\rangle)^2$, while it is suppressed by a further factor $v/\langle\phi\rangle$  in the latter case, due to 
an additional flavon-fermion vertex.
As a consequence, any effect on the properties of the observed Higgs will be suppressed by at least a factor $(v/\langle\phi\rangle)^4$ so that the mixing will affect the $h$ decay widths at most at the level of few percent for $\phi$ satisfying the flavour bounds discussed above.
We leave a more detailed discussion of this interesting aspect of low-energy flavour models to future research.

\section{Conclusions and Outlook}

In this article we discussed the possibility that the DM particle is a singlet under the SM gauge group but carries flavour charges, such that it can communicate with the SM only through flavour gauge bosons or flavour-breaking scalars. We discussed the general features of such flavour portal scenario. We studied an explicit example with an abelian $U(1)$ flavour symmetry. For this model the relevant phenomenological constraints are relic density, direct detection and flavour observables. These measurements often constrain the model to be close to the resonant regime $m_\chi\sim m_{\phi,Z'}/2$, but still future direct detection experiments will be able to probe an important part of the parameter space of the model. 

Extensions of these kinds of models to the leptonic sector are straightforward. Without entering the details, here we just notice that, for the sake of dark matter stability, mixing of the $\chi$ singlets with neutrinos must be avoided, e.g.~by enforcing a conserved symmetry (a natural choice would be the lepton number itself) or by a suitable construction of the UV completion of the model.
While we illustrated our concept for a specific model, 
the basic features of flavour portals models as discussed in Sec.~\ref{sec:gen} are generic. On the other hand, the details of the phenomenology will depend of the specific realisation. Therefore it will be interesting to extend the above discussion to different charges assignments or other flavour groups, in particular non-abelian symmetries. 

{\it Acknowledgments.}--- {\small 
The authors are grateful to the Mainz Institute for Theoretical Physics (MITP) for hospitality and
partial support during the workshop ``Probing the TeV scale and beyond'' where this project was initiated.
We thank P.~Tziveloglou for useful discussions. AC is supported by a Marie Curie Intra-European Fellowship of the European Community's 7th Framework Programme under contract number (PIEF-GA-2012-326948). The work of BZ
is supported by the IISN, an ULB-ARC grant and by the Belgian Federal Science Policy
through the Interuniversity Attraction Pole P7/37.}

\bibliography{BIB_FlavourPortal}
\end{document}